\documentclass[journal,10pt]{IEEEtran}

\usepackage{mathrsfs}
\usepackage{amsfonts,amssymb} 
\usepackage{amsmath,amssymb,amsthm,lipsum}
\usepackage{algorithm}
\usepackage{algorithmic}
\floatname{algorithm}{Network}
\usepackage{graphicx}
\usepackage{bm}
\usepackage{CJK}
\usepackage{enumerate}
\usepackage{diagbox}
\usepackage{float}
\usepackage{caption}
\usepackage{cite}
\usepackage{bbm}
\usepackage{subfigure}
\usepackage{enumitem}
\usepackage{setspace}
\usepackage{bbding}
\usepackage{pifont}
\usepackage{wasysym}
\usepackage{makecell}
\usepackage{titlesec}
\usepackage{color}
\usepackage{xcolor}
\begin{document}

\title{Sufficient Conditions for Unique Optimizer of Two-Dimensional Atomic Norm Minimization Under Multiple Frequencies}

\author{An~Chen,~\IEEEmembership{Member,~IEEE,}
        Wenbo~Xu,~\IEEEmembership{Senior Member,~IEEE,}

\thanks{This work was supported by the National Natural Science Foundation of China (62371053).(\emph{Corresponding author: Wenbo Xu}.)

A. Chen and W. Xu are with the Key Lab of Universal Wireless Communications, Ministry of Education, Beijing University of Posts and Telecommunications, Beijing 100876, China (e-mail: anchen@bupt.edu.cn; xuwb@bupt.edu.cn).

}\vspace{-2pt}}

\maketitle
\begin{abstract}
Atomic norm minimization (ANM) has been extensively applied for gridless angle estimation. However, with the increase of the number of antennas and the communication frequencies in massive MIMO systems, the accompanying beam squint effect significantly degrades angle estimation accuracy. Existing solutions either address this issue only in the one-dimensional (1D) SIMO case, or decouple the two-dimensional (2D) angle estimation into two separate 1D problems, which fails to achieve the optimal solution. In this paper, we employ the multi-frequency model to characterize the beam squint effect in MIMO channels and propose a multi-frequency version of the ANM objective for corresponding 2D angle estimation. To efficiently retrieve the angle parameters, we prove the existence of the equivalent semi-definite program formulation of the ANM objective and develop an algorithm based on the alternating direction method of multipliers for its solutions.  Moreover, we derive the certification conditions of this objective to guarantee the existence of a unique optimal solution.
\end{abstract}


\section{Introduction}

\IEEEPARstart{M}{assive} multiple input multiple output (MIMO)

$Notations$: Boldface letters are used to represent matrices and vectors. Conventional notations $\mathbf{A}^T$, $\mathbf{A}^H$, $\mathbf{A}^{*}$ and $Tr(\mathbf{A})$ respectively represent the transpose, conjugate transpose, conjugate and trace of a matrix $\mathbf{A}$. $\mathbf{0}_{M\times{N}}$ denotes a zero matrix with $M$ rows and $N$ columns.  $\mathbf{A}_{(k,l)}$ is the entry in the $k$-th row and $l$-th column of the matrix $\mathbf{A}$. The $k$-th row and $l$-th column of a matrix are denoted as $\mathbf{A}_{(k,:)}$ and $\mathbf{A}_{(:,l)}$, respectively.  The $l_2$ norm is given by $||\cdot||_2$. $\mathcal{C}_{M} $ represents the universial set $\lbrace 1,\cdots,M \rbrace$. $\mathbbm{R}(\cdot)$ takes the real part of the corresponding element. $\mathbf{I}_{L}$ is an identity matrix of size $L\times L$.  $\mathcal{A} = [\mathbf{A}_{1} \ | \ \cdots \ | \ \mathbf{A}_{P}]$  represents a tensor that consists of $P$ matrices.  For tensors $\mathcal{A} \in \mathbb{C}^{M \times G \times P}$ and $\mathcal{B} \in \mathbb{C}^{M \times N \times P}$, we define the real inner product and complex inner product as $< \mathcal{A},\mathcal{B}>_{\mathbbm{R}}=\sum_{p=1}^P \mathrm{RTr}(\mathbf{A}_p^H\mathbf{B}_p)$ and  $< \mathcal{A},\mathcal{B}> =\sum_{p=1}^P \mathrm{Tr}(\mathbf{A}_p^H\mathbf{B}_p)$, respectively. $\mathrm{RTr}(\cdot)$ denotes the real part of the trace operator. For the matrices $\mathbf{A} \in \mathbb{C}^{N\times L}$ and $\mathbf{B} \in \mathbb{C}^{M\times L}$, $\circledcirc$ is the Khatri–Rao product defined as $\mathbf{A} \circledcirc \mathbf{B}^T = [\mathbf{A}_{(:,1)} \mathbf{B}^T_{(:,1)}  \ |\  \cdots \ |\ \mathbf{A}_{(:,L)} \mathbf{B}^T_{(:,L)}]$.  $\otimes$ and $\odot$ stand for the Kronecker product and the Hadamard product, respectively. $\mathbb{D}_{+}$ represents the set of nonnegative diagonal matrices.

\section{SYSTEM MODEL AND PROBLEM FORMULATION}
\subsection{Received signal}
We consider a 2D multi-frequency model for representing the wideband channel, where $P$ frequency bins in a frequency set $\mathcal{F} = \lbrace f_0,\cdots,Pf_0 \rbrace$ are utilized to characterize the wideband channel. Moreover, we assume that the transmitter (TX) and the receiver (RX) are equipped with an $M$-element uniform linear array (ULA) and an $N_r$-element ULA, respectively. Specifically, the received signal model at the $p$-th frequency can be expressed as follows:
\begin{align}\label{channel response}
 \mathcal{X} &= \sum\limits_{l=1}^{L} [c_l^1 \mathbf{a}(\omega^r_{l},1) \mathbf{b}^T(\omega^t_{l},1) \ | \ \cdots \ | \ c_l^P \mathbf{a}(\omega^r_{l},P) \mathbf{b}^T(\omega^t_{l},P)], \nonumber \\
& = \sum\limits_{l=1}^{L} \mathbf{c}_l \circledcirc \mathbf{A}(\omega^r_{l}) \circledcirc \mathbf{B}^T(\omega^t_{l})
= [ \mathbf{X}_1 \ |\  \cdots \ |\  \mathbf{X}_P],
\end{align}
where  $\mathbf{X}\in \mathbb{C}^{N_r\times N_t \times P} $,  $c_l^p=\beta_l^p s^p$ represents the complex amplitude factor of the $l$-th source at the $p$-th frequency, $\beta^p_l$ and $s^p$ respectively stand for the fading coefficient and the transmitted signal.  $\mathbf{c}_l = [c_l^1, \cdots,c_l^P]$ consists of the amplitude factors of all frequencies. $L$ denotes the number of sources. The $i$-th entries of the array steering vectors $\mathbf{a}(\omega^r_{l},p)$ and $\mathbf{b}(\omega^t_{l},p)$ are respectively defined as follows:
\begin{align}\label{steer vector}
\mathbf{a}(\omega^r_{l},p)_{(i)}= {e}^{-j2\pi (i-1) p\omega_l^r}, i \in \lbrace 0,\cdots,N_r-1\rbrace, \nonumber \\
\mathbf{b}(\omega^t_{l},p)_{(j)}= {e}^{-j2\pi (j-1) p\omega_l^t}, j \in \lbrace 0,\cdots,N_t-1\rbrace,
\end{align}
where $\omega_l^r = \frac{df_0\sin (\theta_{l})}{c}$ and $\omega_l^t = \frac{df_0\sin (\phi_{l})}{c}$ denote the normalized direction of arrival (DOA) and direction of departure (DOD), respectively. $d=\frac{c}{2f_0}$ represents the spacing between two adjacent array element and $c$ denotes the speed of light. The multi-frequency matrces $\mathbf{A}(\omega^r_{l}) = [\mathbf{a}(\omega^r_{l},1), \cdots, \mathbf{a}(\omega^r_{l},P)]$ and $\mathbf{B}(\omega^t_{l}) = [\mathbf{b}(\omega^t_{l},1), \cdots, \mathbf{b}(\omega^t_{l},P)]$.

\subsection{Problem Formulation}
In the noise-free case, to retrieve the DOD and DOAs from the signal matrix $\mathcal{X}$, the atomic norm minimization (ANM) optimization problem is proposed as follows:
\begin{equation}\label{anm unconstrained objective}
\begin{aligned}
 \min_{\mathcal{Y}} || \mathcal{Y} ||_{\mathcal{A}} \ \ \  s.t. \ \ \ \mathcal{Y} = \mathcal{X},
\end{aligned}
\end{equation}
where $\mathcal{A}$ denotes the atomic set, i.e.,
\begin{align}\label{atomic set}
\mathcal{A} = \lbrace  \mathbf{A}(\omega^r) \circledcirc \mathbf{B}^T(\omega^t)\ | \ \omega^r, \omega^t \in [\frac{-\pi}{2},\frac{\pi}{2}], \ ||  \mathbf{c}||_2 \leq 1] \rbrace,
\end{align}
and $|| \mathcal{Y} ||_{\mathcal{A}}$ represents the atomic norm, i.e.,
\begin{align}\label{atomic norm}
|| \mathcal{Y} ||_{\mathcal{A}} = \text{inf} \{ & \sum_{l=1}^L\sum_{p=1}^P |c_{l}^{p}|  \ | \ \mathcal{Y} = \sum_{l}^L \mathbf{c}_{l} \circledcirc \mathbf{A}(\omega^r_l)  \circledcirc \mathbf{B}^T(\omega^t_l), \nonumber \\
& || \mathbf{c}_{l}||_2 \leq 1 \}.
\end{align}

However, directly obtaining DODs and DOAs based on (\ref{anm unconstrained objective}) is difficult. Therefore,  we formulate a dual optimization problem of (\ref{anm unconstrained objective}) for efficient computation
of the atomic decomposition of $\mathcal{Y}$.  The dual problem of (\ref{anm unconstrained objective}) can be expressed as:
\begin{align}\label{dual problem}
\underset{\mathcal{Q} }{\text{max}}  <\mathcal{Q},\mathcal{Y}>_{\mathbbm{R}}, \ \  s.t. \ \ || \mathcal{Q} ||_{\mathcal{A}}^{*} \leq 1,
\end{align}
where $ || \mathcal{Q} ||_{\mathcal{A}}^{*}$ denotes the dual atomic norm, which can be written as:
\begin{align}\label{dual norm trace}
\!\!\!\! || \mathcal{Q} ||_{\mathcal{A}}^{*} &= \underset{ || \mathcal{Y} ||_{\mathcal{A}} \leq 1}{\text{sup}} <\mathcal{Q},\mathcal{Y} >_{\mathbbm{R}}  \nonumber \\
&= \underset{ \omega^t , \omega^r , c^{p}}{\text{sup}} \sum_{p=1}^{P} \mathbbm{R}(\mathbf{b}^T(\omega^t,p)\mathbf{Q}_p^H \mathbf{a}(\omega^r,p) c^{p}) \nonumber \\
&= \underset{ \omega^t , \omega^r , c^{p}}{\text{sup}} \mathbbm{R}(\mathbf{c}^T \bm{\chi}_{\omega^t , \omega^r}) = \underset{ \omega^t , \omega^r , c^{p}}{\text{sup}} |\mathbf{c}^T \bm{\chi}_{\omega^t , \omega^r}| \nonumber \\
& \overset{(\text{a})}= \underset{ \omega^t , \omega^r, c^{p}}{\text{sup}} ||\mathbf{c}||_2 ||\bm{\chi}_{\omega^t , \omega^r}||_2  \overset{(\text{b})}=  \underset{ \omega^t , \omega^r , c^{p}}{\text{sup}} ||\bm{\chi}_{\omega^t , \omega^r}||_2.
\end{align}
The equation (a) follows Cauchy Inequality, and the property of $\mathbf{c}$ in (\ref{atomic norm}) leads to equation (b).  The vector 
$\mathbf{c} = [c^1,\cdots,c^p]^T$ and the dual polynomial vector $\bm{\chi}_{\omega^t , \omega^r} $ is defined as  
\begin{align}\label{dual polynomial vector}
\bm{\chi}_{\omega^t , \omega^r} = & [ \mathbf{b}^T(\omega^t,1)\mathbf{Q}_1^H\mathbf{a}(\omega^r,1),\cdots, \\
& \ \mathbf{b}^T(\omega^t,P)\mathbf{Q}_P^H\mathbf{a}(\omega^r,P)].
\end{align} Since the original convex problem is unconstrained, strong duality naturally holds between (\ref{dual problem}) and (\ref{anm unconstrained objective}), a direct application of strong duality is the certification of the optimality and uniqueness of the solution to (\ref{anm unconstrained objective}) with the help of dual polynomial vectors. 

\section{DUAL CERTIFICATION FOR UNIQUE SOLUTION}
The dual polynomial vector plays an important role in gridless angle estimation, as it paves the way for certifying the optimality. In this section, based on the dual polynomial vector, the conditions required for guaranteeing a unique optimizer to (\ref{anm unconstrained objective}) are derived.

\subsection{Conditions of Dual Polynomial Vector for Unique Optimizer}

\emph {Theorem I:} Given the set of parameters $\mathcal{D} = \lbrace \omega^r_l,\omega^t_l, c_l^p \ | \ l = 1,\cdots,L , p = 1,\cdots,P \rbrace$,   if the following conditions are satisfied, $\mathcal{Y} = \sum_{\omega^r_l,\omega^t_l,c_l^p \in \mathcal{D}} \mathbf{c}_{l} \circledcirc \mathbf{A}(\omega^r_l)  \circledcirc \mathbf{B}^T(\omega^t_l)$ is the unique optimizer to (\ref{anm unconstrained objective}) and is the unique atomic decomposition that fulfills $|| \mathcal{Y} ||_{\mathcal{A}}=\sum_{c_l^p \in \mathcal{D}} |c_l^p|$ :
\begin{enumerate}[leftmargin=10pt, labelsep=6pt,itemindent=8pt,label={(\arabic*)}]
	\item  There exists a dual polynomial vector $\bm{\chi}_{\omega^t , \omega^r} $  that satisfies 
\begin{align}\label{dual polynomial condition i}
\begin{cases}
\bm{\chi}_{\omega^t_l, \omega^r_l} = sign(\mathbf{c}^*_l)=\frac{\mathbf{c}^*_l}{||\mathbf{c}^*_l||_2}, \ \ \forall \omega^t_l, \omega^r_l \in \mathcal{D},\\
|| \bm{\chi}_{\omega^t_l, \omega^r_l} ||_2  \leq 1, \forall \omega^t_l, \omega^r_l \notin \mathcal{D}.
\end{cases} 
\end{align}
\item The elements of the set $\lbrace \mathbf{c}_{l} \circledcirc \mathbf{A}(\omega^r_l)  \circledcirc \mathbf{B}^T(\omega^t_l) \ | \ \omega^r_l,\omega^t_l,c_l^p \in \mathcal{D} \rbrace$ are linearly independent.
\end{enumerate}

\emph {Proof:} \ Firstly, we prove that condition (1) leads to $|| \mathcal{Y} ||_{\mathcal{A}}=\sum_{c_l^p \in \mathcal{D}} |c_l^p|$. 
\begin{align}
||\mathcal{Y}||_{\mathcal{A}} \ \overset{(a)} \geq \ ||\mathcal{Y}||_{\mathcal{A}} || \mathcal{Q} ||_{\mathcal{A}}^{*}  \ \overset{(b)}\geq \  <\mathcal{Q},\mathcal{Y}>_{\mathbbm{R}},
\end{align}
where (a) holds since $|| \mathcal{Q} ||_{\mathcal{A}}^{*} = \underset{ \omega^t , \omega^r , c_l^{p}}{\text{sup}} ||\bm{\chi}_{\omega^t , \omega^r}||_2 \leq 1$
according to condition (1), the inequality (b) follows Hölder’s inequality. Then, by taking the decomposition into account, we get
\begin{align}\label{proof condition i}
& \! \! \! \! \! \! \! <\mathcal{Q},\mathcal{Y}>_{\mathbbm{R}} \nonumber \\
& \! \! = \mathbbm{R}( Tr(\sum_{p=1}^{P} \mathbf{Q}_p^H  \sum_{\omega^r_l,\omega^t_l,c_l^p \in \mathcal{D}} c^{p}_l\mathbf{a}(\omega^r_l,p)\mathbf{b}^T(\omega^t_l,p))) \nonumber \\
& \! \!  = \mathbbm{R}( \sum_{p=1}^{P}\sum_{\omega^r_l,\omega^t_l,c_l^p \in \mathcal{D}} c^{p}_l \mathbf{b}^T(\omega^t_l,p)\mathbf{Q}_p^H \mathbf{a}(\omega^r_l,p)) \nonumber \\
& \! \! \overset{(a)}= \sum_{c_l^p \in \mathcal{D}} \mathbf{c}_l^T sign(\mathbf{c}_l^*)= \sum_{c_l^p \in \mathcal{D}} |c_l^p| \overset{(b)}\geq ||\mathcal{Y}||_{\mathcal{A}},
\end{align}
where (a) holds because of condition (1), and (b) follows the definition of atomic norm in (\ref{atomic norm}). According to (\ref{proof condition i}), we conclude that  $||\mathcal{Y}||_{\mathcal{A}} = <\mathcal{Q},\mathcal{Y}>_{\mathbbm{R}} = \sum_{c_l^p \in \mathcal{D}} |c_l^p|$.

Next, we prove the uniqueness of the decomposition. If there exists another decomposition $\mathcal{Y} = \sum_{\omega^r_{l^{\prime}},\omega^t_{l^{\prime}},c_{l^{\prime}}^p \in \mathcal{D}^{\prime}} \mathbf{c}_{l^{\prime}} \circledcirc \mathbf{A}(\omega^r_{l^{\prime}})  \circledcirc \mathbf{B}^T(\omega^t_{l^{\prime}}) $ that satifies $||\mathcal{Y}||_{\mathcal{A}} = \sum_{c_{l^{\prime}}^p \in \mathcal{D}^{\prime}}  |c_{l^{\prime}}^p| $, $\mathcal{D}^{\prime}$ must contain $\omega^r_{l^{\prime}} \notin \mathcal{D} $ due to the condition  (2) in Theorem I. Therefore, we obtain 
\begin{align}\label{proof condition ii}
& \! \! \! \! <\mathcal{Q},\mathcal{Y}>_{\mathbbm{R}}   \nonumber \\
& =  \mathbbm{R}( \sum_{\omega^r_{l^{\prime}},\omega^t_{l^{\prime}},c_{l^{\prime}}^p \in \mathcal{D}} \mathbf{c}_{l^{\prime}}^T \bm{\chi}_{\omega^t_{l^{\prime}}, \omega^r_{l^{\prime}}}  +   \sum_{\omega^r_{l^{\prime}},\omega^t_{l^{\prime}},c_{l^{\prime}}^p \notin \mathcal{D}} \mathbf{c}_l^T \bm{\chi}_{\omega^t , \omega^r} ) \nonumber \\ 
&  \overset{(a)} < \sum_{c_{l^{\prime}}^p \in \mathcal{D}}  |c_{l^{\prime}}^p| + \sum_{c_{l^{\prime}}^p \notin \mathcal{D}}  |c_{l^{\prime}}^p| = ||\mathcal{Y}||_{\mathcal{A}},
\end{align}
where (a) holds because condition (1) implies the inequality $ \sum_{\omega^r_{l^{\prime}},\omega^t_{l^{\prime}},c_{l^{\prime}}^p \notin \mathcal{D}^{\prime}} \mathbf{c}_{l^{\prime}}^T \bm{\chi}_{\omega^t_{l^{\prime}}, \omega^r_{l^{\prime}}}< \sum_{\omega^r_{l^{\prime}},\omega^t_{l^{\prime}},c_{l^{\prime}}^p\notin \mathcal{D}^{\prime}} ||\mathbf{c}_l||_2 ||\bm{\chi}_{\omega^r_{l^{\prime}},\omega^t_{l^{\prime}}} ||_2$ $  < \sum_{c_{l^{\prime}}^p\notin \mathcal{D}^{\prime}} ||\mathbf{c}_l||_2  \leq \sum_{c_{l^{\prime}}^p \notin \mathcal{D}} |c_l^p| $. The inequality (\ref{proof condition ii}) shows that the parameters outside of $\mathcal{D}$ contradicts the strong duality between primal problems (\ref{anm unconstrained objective}) and dual problems (\ref{dual problem}), which implies that another decomposition $\mathcal{Y} = \sum_{\omega^r_{l^{\prime}},\omega^t_{l^{\prime}},c_{l^{\prime}}^p \in \mathcal{D}^{\prime}} \mathbf{c}_{l^{\prime}} \circledcirc \mathbf{A}(\omega^r_{l^{\prime}})  \circledcirc \mathbf{B}^T(\omega^t_{l^{\prime}}) $ can not exist.

\subsection{Dual Polynomial Vector Construction}
As the sufficient condition for the existence of a unique optimizer is given in Theorem I, a question naturally arises: whether there exists a dual polynomial vector $\bm{\chi}_{\omega^t , \omega^r} $ that satisfies Theorem I? The following theorem states an affirmative answer and lists the conditions that guarantee the existence of such a dual polynomial vector.

\emph {Theorem II:} If $\Delta^p(\bm{\bar{\omega}}) >\frac{1.19}{\min \lbrace \lfloor \frac{N_r-1}{4} \rfloor,\lfloor \frac{N_t-1}{4} \rfloor \rbrace}, \forall p \in \mathcal{C}_P $, $G =\max \lbrace \lfloor \frac{N_r-1}{2} \rfloor,\lfloor \frac{N_t-1}{2} \rfloor \rbrace \geq 512$, and $|sign(\mathbf{c}^*_{l,(p)})| \leq \frac{1}{\sqrt{P}}, \forall l \in \mathcal{C}_L, \forall p \in \mathcal{C}_P$, then there exists a dual polynomial vector that satisfies condition (1) in Theorem I.

\emph {Proof:} The proof is divided into five parts. For the first part, to utilize the results established in \cite{1} for the subsequent proof, the polynomial vector $\bm{\chi}_{\omega^t_l, \omega^r_l}$ is modified to obtain its equivalent shifted-coordinate representation $\tilde{\bm{\chi}}_{\omega^t_l, \omega^r_l}$. Next, based on the Fejér kernel, a set of interpolation coefficients is designed to construct $\tilde{\bm{\chi}}_{\omega^t_l, \omega^r_l}$, then we demonstrated that for the actual angle parameters $\omega^t_l, \omega^t_l, c^p_l \in \mathcal{D}$, the constructed polynomial vector satisfies $\tilde{\bm{\chi}}_{\omega^t_l, \omega^r_l} =  sign(\mathbf{c}^{*}_l)$. Then, the upper bounds on the $\ell_\infty$-norm of the designed interpolation coefficients are derived for any impractical parameter point $\omega^t_l, \omega^r_l \notin \mathcal{D}$, and it is proved that $ ||\bm{\chi}_{\omega^t_l, \omega^r_l}||_2 \leq 1$ holds for any impractical point,. Finally, based on the above steps, sufficient conditions for the existence of $\bm{\chi}_{\omega^t_l, \omega^r_l}$ are established.

To begin with, we present an alternative version of  $\mathbf{b}(\omega^t,p)$ and $\mathbf{a}(\omega^r,p)$ with shifted coordinates, i.e.,  
\begin{align}\label{steer vector}
\tilde{\mathbf{a}}(\omega^r_{l},p)_{(i)} & = {e}^{-j2\pi i p\omega_l^r}, i \in \lbrace -2N_a,\cdots,2N_a\rbrace, \nonumber \\
\tilde{\mathbf{b}}(\omega^t_{l},p)_{(i)} & = {e}^{-j2\pi i p\omega_l^t}, i \in \lbrace -2M_a,\cdots,2M_a \rbrace,
\end{align}
where $N_a = \frac{N_r-1}{4}$ and $M_a = \frac{N_t-1}{4}$. Then we can construct a similar formulation of condition (1) in Theorem I, i.e.,
\begin{align}\label{dual polynomial condition i shifted}
\begin{cases}
\tilde{\bm{\chi}}_{\omega^t_l, \omega^r_l} =\mathbf{d}^*_l, \ \ \forall \omega^t_l, \omega^r_l \in \mathcal{D},\\
|| \tilde{\bm{\chi}}_{\omega^t_l, \omega^r_l} ||_2  \leq 1, \forall \omega^t_l, \omega^r_l \notin \mathcal{D}.
\end{cases}  
\end{align}
where the $p$-th entry of $\mathbf{d}^*_l$ is defined as $ \mathbf{d}^{*}_{l,(p)} = sign(\mathbf{c}^{*}_{l,(p)})e^{j\pi(N_r-1)\omega_l^r p}e^{j\pi(N_t-1)\omega_l^t p}$.

Note that the condition (\ref{dual polynomial condition i}) is satisfied as long as the condition (\ref{dual polynomial condition i shifted}) is satisfied. Because $||\tilde{\bm{\chi}}_{\omega^t , \omega^r} ||_2 = ||\bm{\chi}_{\omega^t , \omega^r} ||_2$ and the equation in (\ref{dual polynomial condition i shifted}) can be simply transformed to (\ref{dual polynomial condition i}) by shifting coordinates, i.e., the
$p$-th entry $\bm{\chi}_{\omega^t_l, \omega^r_l,(p)} = \tilde{\bm{\chi}}_{\omega^t_l, \omega^r_l,(p)} e^{-j\pi(N_r-1)\omega_l^r p}e^{-j\pi(N_t-1)\omega_l^t p} = \mathbf{d}^{*}_{l,(p)} e^{-j\pi(N_r-1)\omega_l^r p}e^{-j\pi(N_t-1)\omega_l^t p}$. 

Secondly, according to the above definition, the proof of the existence of $\bm{\chi}_{\omega^t , \omega^r}$ is transformed to the proof of the existence of $\tilde{\bm{\chi}}_{\omega^t , \omega^r} $. Therefore, we construct the dual polynomial  $\tilde{\bm{\chi}}_{\omega^t , \omega^r} $ (\ref{dual polynomial vector fkernel}) in the next page,
\begin{figure*}
\begin{align}\label{dual polynomial vector fkernel}
\tilde{\bm{\chi}}_{\omega^t , \omega^r}  &= 
\begin{bmatrix} \sum_{l=1}^L \bm{\alpha}_{1,(l)}K_1^{(0,0)}(\bm{\omega}-\bm{\omega}_l) + \bm{\beta}_{1,(l)}K_1^{(1,0)}(\bm{\omega}-\bm{\omega}_l) + \bm{\epsilon}_{1,(l)}K_1^{(0,1)}(\bm{\omega}-\bm{\omega}_l) \\ \cdots \\ \sum_{l=1}^L \bm{\alpha}_{P,(l)}K_P^{(0,0)}(\bm{\omega}-\bm{\omega}_l) + \bm{\beta}_{P,(l)}K_P^{(1,0)}(\bm{\omega}-\bm{\omega}_l) + \bm{\epsilon}_{P,(l)}K_P^{(0,1)}(\bm{\omega}-\bm{\omega}_l)
\end{bmatrix}.
\end{align}
\end{figure*}
where $K_p^{(i,j)}(\bm{\omega})=\frac{\partial^i \partial^j K_p(\bm{\omega})}{\partial (\omega^r)^i \partial (\omega^t)^j }$ denotes the second order partial derivative.  $\bm{\alpha}_{p,(l)}$, $\bm{\beta}_{p,(l)}$ and $\bm{\epsilon}_{p,(l)}$ are the interpolation coefficients. In addition, we assume $\omega_l^r,\omega_l^t \in [0,1]$ instead of $[-0.5,0.5]$ because they are equivalent due to the periodicity of the following Fejér kernel.
\begin{align}\label{fkernel}
K_p(\bm{\omega}) & = K_p(\omega^r)K_p(\omega^t), \bm{\omega} = [ \omega^t,\omega^r  ], \\
K_p(\omega^r)& =\frac{1}{pN_a}\sum_{k=-2N_a}^{2N_a}g_{N_a}(k)e^{-j2\pi k \omega^r p}, \\
K_p(\omega^t)& =\frac{1}{pM_a}\sum_{k=-2M_a}^{2M_a}g_{M_a}(k)e^{-j2\pi k \omega^t p}, \\
g_{N}(k) &= \frac{1}{N}\sum_{t=max(k-N,-N)}^{min(k+N,N)}(1-\frac{|t|}{N})(1-\frac{|k-t|}{N}).
\end{align}

In order for the interpolation coefficients at the $p$-th subcarrier to satisfy the inequality in (\ref{dual polynomial condition i shifted}),  we must have the partial derivates as $0$, i.e.,
\begin{align}\label{polynomial equation II}
\tilde{\bm{\chi}}_{\omega^t , \omega^r,(p)}^{(0,1)} = 0, \ \tilde{\bm{\chi}}_{\omega^t , \omega^r,(p)}^{(1,0)} = 0, \ \ \forall \omega^t_l, \omega^r_l \notin \mathcal{D}.
\end{align}
Therefore, combing (\ref{polynomial equation II}) and the equation in (\ref{dual polynomial condition i shifted}). The interpolation coefficients can be computed by solving the following equation:
\begin{align}\label{polynomial equation solution}
\begin{bmatrix} \mathbf{E}^{(0,0)}_{p}  \mathbf{E}^{(1,0)}_{p} \mathbf{E}^{(0,1)}_{p} \\ 
\mathbf{E}^{(1,0)}_{p}  \mathbf{E}^{(2,0)}_{p} \mathbf{E}^{(1,1)}_{p} \\
\mathbf{E}^{(0,1)}_{p}  \mathbf{E}^{(1,1)}_{p} \mathbf{E}^{(0,2)}_{p}
\end{bmatrix}
\begin{bmatrix}
\bm{\alpha}_p \\
\bm{\beta}_p \\
\bm{\epsilon}_p 
\end{bmatrix} = \mathbf{E}_p
\begin{bmatrix}
\bm{\alpha}_p \\
\bm{\beta}_p \\
\bm{\epsilon}_p 
\end{bmatrix} =
\begin{bmatrix}
 \mathbf{d}^{*}_{1,(p)} \\
 \cdots \\
 \mathbf{d}^{*}_{L,(p)} \\
 \mathbf{0}_{L\times 1} \\
 \mathbf{0}_{L\times 1}
\end{bmatrix},
\end{align}
where the $(m,n)$-th entry $\mathbf{E}^{(i,j)}_{p,(m,n)} = K_p^{(i,j)}(\bm{\omega}_m - \bm{\omega}_n),  \forall i,j \in \lbrace 0,1,2 \rbrace, \ m,n \in \lbrace 1,\cdots,L \rbrace$. $\bm{\alpha}_{p} = [\bm{\alpha}_{p,(1)},\cdots,\bm{\alpha}_{p,(L)}]\in \mathbb{C}^{L}$, $\bm{\beta}_{p} = [\bm{\beta}_{p,(1)},\cdots,\bm{\beta}_{p,(L)}]\in \mathbb{C}^{L}$ and $\bm{\epsilon}_{p} = [\bm{\epsilon}_{p,(1)},\cdots,\bm{\epsilon}_{p,(L)}]\in \mathbb{C}^{L}$. $\mathbf{0}_{L\times 1}$ is a zero vector of size $L$. The solution to (\ref{polynomial equation solution}) can be identified by the inverse of $\mathbf{E}_p$, i.e.,
\begin{align}\label{polynomial coeff}
\begin{bmatrix}
\bm{\alpha}_p \\
\bm{\beta}_p \\
\bm{\epsilon}_p 
\end{bmatrix} = \mathbf{E}_p^{-1}
\begin{bmatrix}
 \mathbf{d}^{*}_{l,(p)} \\
 \cdots \\
 \mathbf{d}^{*}_{L,(p)} \\
 \mathbf{0}_{L\times 1} \\
 \mathbf{0}_{L\times 1}
\end{bmatrix},
\end{align}
where $\mathbf{E}_p^{-1}$ can be rewritten based on Schur complement \cite{1} as
\begin{align}
\mathbf{E}_p^{-1} = 
\begin{bmatrix} 
\mathbf{I}_{L\times L} \mathbf{S}^{-1}_3\\
-\mathbf{S}_1^{-1} \mathbf{S}_2 \mathbf{S}^{-1}_3\\
(\mathbf{E}_p^{(0,2)})^{-1} (\mathbf{E}_p^{(1,1)}\mathbf{S}_1^{-1} \mathbf{S}_2 - \mathbf{E}_p^{(0,1)})\mathbf{S}^{-1}_3
\end{bmatrix},
\end{align}
where $\mathbf{S}_1 = \mathbf{E}^{(2,0)}_p - \mathbf{E}^{(1,1)}_p (\mathbf{E}^{(0,2)}_p)^{-1}\mathbf{E}^{(1,1)}_p $, $\mathbf{S}_2 = \mathbf{E}^{(1,0)}_p - \mathbf{E}^{(1,1)}_p (\mathbf{E}^{(0,2)}_p)^{-1} \mathbf{E}^{(0,1)}_p$ and $\mathbf{S}_3 = \mathbf{E}^{(0,0)}_p + \mathbf{S}_2^T\mathbf{S}_1^{-1}\mathbf{S}_2 - \mathbf{E}^{(0,1)}_p(\mathbf{E}^{(0,2)}_p)^{-1}  \mathbf{E}^{(0,1)}_p$.  
Notice that the invertibility of $\mathbf{E}_p$ is essential for the construction of $\tilde{\bm{\chi}}_{\omega^t , \omega^r}$, a proof of its invertibility is provided in Appendix A. 

Thirdly, some preliminary definitions and assumptions are given in this part to show the upper bounds of the interpolation coefficients.

Define the separation of $\omega^r$ under the $p$-th frequency based on the closest wrap-around distance between two distinct angles, i.e.,
\begin{align}\label{separation for r}
& \Delta^p(\omega^r) = \underset{m\neq n}{\min} \lbrace  p|\omega_m^r - \omega_n^r| \ mod \ 1,\nonumber \\
& \ \ \ \ \ \ \ \ \ \ \ \ \ \ \ \ \ \ \ 1 - (p|\omega_m^r -  \omega_n^r|  \ mod \ 1)\rbrace,
\end{align}
this definition also holds for $\omega^t$ under the $p$-th frequency. Note that as $p|\omega_m^r - \omega_n^r|, \ p \ge 2$ could be larger than $1$, we can only keep its fractional part to calculate the separation due to the periodicity of Fejér kernel.

Moreover, define the separation of $\bm{\omega} = \lbrace \omega^t, \omega^r \rbrace$ under the $p$-th frequency as follows:
\begin{align}\label{separation for omega}
\Delta^p(\bm{\omega})=\underset{m\neq n}{\min} \max \lbrace \Delta^p(\omega^r), \Delta^p(\omega^t)\rbrace.
\end{align}

\emph {Lemma I:} If $\Delta^p(\bm{\omega}) >\frac{1.19}{\min \lbrace M_a,N_a \rbrace}$ and $f_c = \max \lbrace 2M_a,2N_a \rbrace \geq 512$,  and the amplitude $|\mathbf{d}^{*}_{l,(p)}| = |sign(\mathbf{c}^{*}_{l,(p)})| \leq \frac{1}{\sqrt{P}}, \forall l \in \mathcal{C}_L$, then we have
\begin{align}\label{coeff upper bound}
&||\bm{\alpha}_p||_{\infty} \leq \frac{1.0533p^2}{\sqrt{P}}, ||\bm{\beta}_p||_{\infty} \leq \frac{0.9556\times 10^{-2}p}{\sqrt{P}f_c^3}, \nonumber \\
&||\bm{\epsilon}_p||_{\infty}\leq  \frac{2.7650\times 10^{-2}p}{\sqrt{P}f_c} \ \text{and} \ \alpha_{p,(1)} \geq (2+1.0533p^2).
\end{align}
\emph {Proof:} See Appendix B.

\emph {Lemma II:} Under the hypotheses of Lemma I and assuming $c_{\alpha}=1.0533$, $c_{\beta}=0.9556\times 10^{-2}$ and $c_{\epsilon}=2.7650\times 10^{-2}$, we obtain $|| \tilde{\bm{\chi}}_{\omega^t, \omega^r} ||_2  \leq 1, \forall \bm{\omega} = [ \omega^t, \omega^r ] \notin \mathcal{D}.$ 

\emph {Proof:} According to the definition of $\tilde{\bm{\chi}}_{\omega^t, \omega^r}$, as long as $|\tilde{\bm{\chi}}_{\omega^t, \omega^r,(p)}|\le \frac{1}{\sqrt{P}},\forall p \in \mathcal{C}_P$, we have $|| \tilde{\bm{\chi}}_{\omega^t_l, \omega^r_l} ||_2 < 1$. Therefore, we begin by proving $|\tilde{\bm{\chi}}_{\omega^t, \omega^r,(p)}|\le \frac{1}{\sqrt{P}},\forall p \in \mathcal{C}_P$.
\begin{align}\label{polynomial modulus p}
& |\tilde{\bm{\chi}}_{\omega^t, \omega^r,(p)}|  \nonumber \\
& = \sum_{l=1}^L \bm{\alpha}_{p,(l)}K_p^{(0,0)}(\bm{\omega}-\bm{\omega}_l) + \bm{\beta}_{p,(l)}K_p^{(1,0)}(\bm{\omega}-\bm{\omega}_l) \nonumber \\
&\ \ \ \ \ \ \ \ + \bm{\epsilon}_{p,(l)}K_p^{(0,1)}(\bm{\omega}-\bm{\omega}_l) \nonumber \\
& \leq ||\bm{\alpha}_{p}||_{\infty} \sum_{l=1}^L |K_p^{(0,0)}(\bm{\omega}-\bm{\omega}_l)| + ||\bm{\beta}_{p}||_{\infty} \nonumber \\
& \ \ \ \ \sum_{l=1}^L |K_p^{(1,0)}(\bm{\omega}-\bm{\omega}_l)| +  ||\bm{\epsilon}_{p}||_{\infty}\sum_{l=1}^L |K_p^{(0,1)}(\bm{\omega}-\bm{\omega}_l)| \nonumber \\
& \leq \frac{c_\alpha}{\sqrt{P}}\sum_{l=1}^L |K_1^{(0,0)}(p(\bm{\omega}-\bm{\omega}_l) \ mod \ 1)| + \frac{c_\beta}{\sqrt{P}f_c^3} \nonumber \\
& \ \ \ \ \sum_{l=1}^L |K_1^{(1,0)}(p(\bm{\omega}-\bm{\omega}_l) \ mod \ 1)  + \frac{c_\epsilon}{\sqrt{P}f_c} \nonumber \\
& \ \ \ \  \sum_{l=1}^L | K_1^{(0,1)}(p(\bm{\omega}-\bm{\omega}_l) \ mod \ 1)|.
\end{align}
Moreover, by defining $\mathcal{D}_{near} = \cup_{l=1}^L \lbrace \bm{\omega}: 0<\max \lbrace \bm{\omega} - \bm{\omega}_l \rbrace \leq 0.1224/\max \lbrace M_a,N_a \rbrace \rbrace $ and $\mathcal{D}_{far} = [0,1)\times [0,1) \backslash \mathcal{D}_{near} $,  it has been shown in \cite{3} that for $\bm{\omega} \in \mathcal{D}_{far}$ and $p=1$, 
\begin{align}\label{far less one}
& c_\alpha\sum_{l=1}^L |K_1^{(0,0)}(\bm{\omega}-\bm{\omega}_l )| + \frac{c_\beta}{f_c^3}\sum_{l=1}^L |K_1^{(1,0)}( \bm{\omega}-\bm{\omega}_l)| + \frac{c_\epsilon}{f_c} \nonumber \\
&  \ \ \ \sum_{l=1}^L |K_1^{(0,1)}(\bm{\omega}-\bm{\omega}_l)| < 1.
\end{align}
Therefore, we have $|\tilde{\bm{\chi}}_{\omega^t, \omega^r,(1)}|\le \frac{1}{\sqrt{P}}$. Meanwhile, by comparing the upper bound of $|\tilde{\bm{\chi}}_{\omega^t, \omega^r,(p)}|, p>1$ with $|\tilde{\bm{\chi}}_{\omega^t, \omega^r,(1)}|$, we notice that the only difference lies in the term $p(\bm{\omega}-\bm{\omega}_l) \ mod \ 1$, which can not affect the upper bound of $K_1^{(0,1)}(p(\bm{\omega}-\bm{\omega}_l)), K_1^{(1,0)}(p(\bm{\omega}-\bm{\omega}_l)) \ \text{and} \ K_1^{(0,1)}(p(\bm{\omega}-\bm{\omega}_l))$ according to \cite{1}. Thus, we conclude that $||\tilde{\bm{\chi}}_{\omega^t, \omega^r}||_2 \le 1$ when $\bm{\omega} \in \mathcal{D}_{far}$.

On the other hand, according to the paragraph mentioned before, if $\mathbf{E}$ is invertible, then we have 
\begin{align}
& \tilde{\bm{\chi}}_{\omega^t , \omega^r,(p)}^{(0,1)} = 0, \ \tilde{\bm{\chi}}_{\omega^t , \omega^r,(p)}^{(1,0)} = 0, \\ \nonumber
& |\tilde{\bm{\chi}}_{\omega^t, \omega^r,(p)}|=1, \forall p \in \mathcal{C}_P, \forall \bm{\omega}= \lbrace \omega^t, \omega^r \rbrace \in \mathcal{D}.
\end{align}
Therefore, if we prove the Hessian matrix of $\tilde{\bm{\chi}}_{\omega^t , \omega^r,(p)}$ is negative definite, i.e.,
\begin{align}\label{hessian matrix}
\mathbf{H} = 
\begin{bmatrix}    
\tilde{\bm{\chi}}_{\omega^t , \omega^r,(p)}^{(2,0)}  \ \ \tilde{\bm{\chi}}_{\omega^t , \omega^r,(p)}^{(1,1)} \\
\tilde{\bm{\chi}}_{\omega^t , \omega^r,(p)}^{(1,1)}  \ \ \tilde{\bm{\chi}}_{\omega^t , \omega^r,(p)}^{(0,2)}
\end{bmatrix} \prec 0,
\end{align}
then local concavity would imply $|\tilde{\bm{\chi}}_{\omega^t , \omega^r,(p)}|<1,  \forall \bm{\omega} \in \mathcal{D}_{near}$.

According to \cite{1}, it can be concluded that
\begin{align}\label{derivate bound}
\tilde{\bm{\chi}}_{\omega^t , \omega^r,(p)}^{(2,0)} \leq -1.7855p^2f_c^2 -4.194f_c^2, \nonumber \\
\tilde{\bm{\chi}}_{\omega^t , \omega^r,(p)}^{(1,1)} \leq 0.6879p^2f_c + 0.1711p^2f_c^2,
\end{align}
which also holds for $\tilde{\bm{\chi}}_{\omega^t , \omega^r,(p)}^{(2,0)}$. Since $det(H)=|\tilde{\bm{\chi}}_{\omega^t , \omega^r,(p)}^{(2,0)}||\tilde{\bm{\chi}}_{\omega^t , \omega^r,(p)}^{(0,2)}| - |\tilde{\bm{\chi}}_{\omega^t , \omega^r,(p)}^{(1,1)}|^2>0$, $Tr(H)=|\tilde{\bm{\chi}}_{\omega^t , \omega^r,(p)}^{(2,0)}| + |\tilde{\bm{\chi}}_{\omega^t , \omega^r,(p)}^{(0,2)}|<0$, the Hessian matrix $\mathbf{H}$ is negative definite  and thereby $||\tilde{\bm{\chi}}_{\omega^t, \omega^r}||_2 \le 1$ when $\bm{\omega} \in \mathcal{D}_{near}$.

\section{SDP FORMULATION OF ANM OBJECTIVE}
The previous section provides a sufficient condition for the uniqueness of the solution to (\ref{anm unconstrained objective}), but an algorithm for obtaining the solution is still absent. In this section, we first present an equivalent and tractable SDP formulation of (\ref{anm unconstrained objective}), and then design an algorithm based on the alternating direction method of multipliers (ADMM) for obtaining its optimal solution.

\subsection{Preliminaries}
\emph{Definition 1}: The matrix satisfying the following form is termed as an irregular Vandermonde matrix \cite{6,7}, i.e.,
\begin{align}\label{irr vand}
\mathbf{W}(\bm{\theta}) & = [\mathbf{z}^{r_{1}}(\bm{\theta}),\cdots,\mathbf{z}^{r_{N}}(\bm{\theta})]^{T}  = [\mathbf{w}(\theta_{1}),\cdots,\mathbf{w}(\theta_{L})],
\end{align}
with
\begin{align}\label{irr vand com}
\mathbf{z}(\bm{\theta}) = [e^{-j\pi sin(\bm{\theta}_{1})},\cdots,e^{-j\pi sin(\bm{\theta}_{L})}]^T \in \mathbb{C}^{L}
\end{align}
where $\mathbf{r}= [r_1,\cdots,r_N] \in \mathbb{R}^{N}$ denotes the indices of the $N$ antenna elements, $\bm{\theta} = [\theta_{1},\cdots,\theta_{L}] \in \mathbb{C}^{L}$ consists of angle parameters corresponds to $L$ paths. For the regular Vandermonde matrix, $\mathbf{r}=[0,\cdots,N-1]$. In contrast, the irregularity is reflected in the non-linear distribution of elements in $\mathbf{r}$. 

\emph {Lemma III (Corollary 4.27 in \cite{8}): For the multivariate polynomial $T_{\mathbf{Q}}(\mathbf{x},\mathbf{w}) = \mathbf{x}^H\mathbf{Q}\mathbf{w}$ with irregular Vandermonde vector $\mathbf{x}\in \mathbb{C}^M$ and $\mathbf{w} \in \mathbb{C}^N$, the necessary and sufficient condition for $|T_{\mathbf{Q}}(\mathbf{x},\mathbf{w})| < \gamma$ is}
\begin{align}
\begin{bmatrix} \gamma^2 \mathbf{P} & vec(\mathbf{Q})  \\ vec(\mathbf{Q})^H & 1 \end{bmatrix} \succeq 0,
\end{align}
with
\begin{align}\label{P matrix} 
&\mathbf{P} \succeq 0 \in \mathbb{C}^{MN\times MN},  \sum_{i=1}^{MN-k} \mathbf{P}_{(i,i+k)} = \delta(k), \nonumber \\
& (\mathbf{x} \otimes \mathbf{w})^H  \mathbf{P} (\mathbf{x} \otimes \mathbf{w}) = 1.
\end{align}

\subsection{Construction of SDP Formulation}

The ANM objective (\ref{anm unconstrained objective})  is rendered a computationally intractable semi-infinite programming problem, because the atoms  $\mathbf{A}(\omega^r) \circledcirc \mathbf{B}^T(\omega^t)$  are defined over a continuous domain, requiring the optimal atoms to be selected from an infinite number of candidates. Therefore, based on the preliminaries, we present its equivalent and tractable SDP formulation for obtaining the optimal solution.

\emph{Proposition I:} If the constraint in (\ref{dual problem}) is satisfied by $|\mathbf{b}^T(\phi,p)\mathbf{Q}_p^H \mathbf{a}(\theta,p)|<1/\sqrt{P}, p\in \mathcal{C}_P$, then the dual problem (\ref{dual problem}) is equivalent to the following dual SDP formulation.
\begin{align}\label{dual SDP}
&\underset{\mathcal{Q},\mathbf{P}^r,\mathbf{P}^t }{\text{max}}  <\mathcal{Q},\mathcal{Y}>_{\mathbb{R}},  \ \ \ {\rm s.t.} \ \ \mathbf{G}_p  = \begin{bmatrix} \mathbf{P}^r / P  & \mathbf{\bar{Q}}_p  \\ \mathbf{\bar{Q}}_p^H & \mathbf{P}^t \end{bmatrix} \succeq 0, \nonumber \\
& \ \ \ \mathbf{\bar{Q}}_p = R(\mathbf{Q}_p), p \in \mathcal{C}_P,  \sum_{i=1}^{N_R-k} \mathbf{P}^r_{(i,i+k)} = \delta(k), \nonumber \\
& \ \ \ \sum_{i=1}^{N_T-k} \mathbf{P}^t_{(i,i+k)} = \delta(k), \mathbf{P}^r \succeq 0, \ \mathbf{P}^t \succeq 0,
\end{align}
where $\mathcal{Q} = [\mathbf{Q}_1 \ |\ \cdots \ | \ \mathbf{Q}_P]\in \mathbb{C}^{N_r \times N_t \times P}$  is the dual variable corresponds to $P$ subcarriers. Moreover, by defining $N_R = P(N_r - 1) + 1$ and $N_T = P(N_t - 1) + 1$, $R(\cdot)$ denotes the mapping from size $N_r \times N_t$ to $N_R \times N_T$, i.e.,
\begin{equation}\label{R trans}
\begin{aligned}
\! & R(\mathbf{Q}_p)_{i,j} = \\
&\begin{cases}
\mathbf{Q}_{p,(r,t)}, & \makecell[lc]{(i,j) = ((r - 1)p + 1,(t - 1) p + 1),}  \\
\mathbf{Q}_{p,(1,1)}, & (i,j) = (1,1). \\
\end{cases}
\end{aligned} 
\end{equation}
The conjugate operator $R^{*}(\cdot)$ describes the adjoint mapping of $R(\cdot)$, i.e., the mapping from size $N_R \times N_T$ to $N_r \times N_t$.

\emph{Proof:} See Appendix C.

Similar to Lemma $I$, the above proposition relies on the assumption $|\mathbf{b}^T(\omega^t,p)\mathbf{Q}_p^H \mathbf{a}(\omega^r,p)| = |\bm{\chi}_{\omega^t, \omega^r,(p)} | < \frac{1}{\sqrt{P}}$, where the element $\bm{\chi}_{\omega^t, \omega^r,(p)} $ in the dual polynomial vector is required to be less than $\frac{1}{\sqrt{P}}$. Intuitively, this constraint ensures that no single frequency component exceeds the threshold $\frac{1}{\sqrt{P}}$ and dominates the multi-frequency signal, which would otherwise degrade the signal into a single-frequency representation and compromise angular estimation accuracy \cite{2}. This assumption may not be necessary in practice, and simulation results are provided in the next section to show that angle parameters can be accurately estimated even without it.

However, the relationship between the variables $\mathcal{Q},\mathbf{P}^r,\mathbf{P}^t$ and the angular parameters $\omega^t, \omega^r$ is difficult to comprehend. Therefore, following the results presented in \cite{2,4,6}, we consider solving the dual problem of (\ref{dual SDP}), which corresponds to the SDP formulation of the primal problem (\ref{anm unconstrained objective}).

\emph{Proposition II:} The dual problem of (\ref{dual SDP}) admits the following expression:
\begin{align}\label{primal SDP} 
 & \underset{\mathbf{T}^r,\mathbf{T}^t}{\text{min}} \ \ \ \frac{1}{N_R} Tr(\mathbf{T}^r) + \frac{1}{N_T} Tr(\mathbf{T}^t) \nonumber \\
{\rm s.t.} & \ \ \ \mathbf{Z}_p = \begin{bmatrix}2\mathbf{T}^r  & \mathbf{\bar{Y}}_p  \\ \mathbf{\bar{Y}}_p^H & 2\mathbf{T}^t/P \end{bmatrix} \succeq 0, \mathbf{\bar{Y}}_p = R(\mathbf{Y}_p), p \in \mathcal{C}_P,
\end{align}
where $\mathbf{T}^r$ and $\mathbf{T}^t$ can be respectively represented by the irregular Vandermonde matrices $\mathbf{W}(\bm{\omega}^r)$ and $\mathbf{W}(\bm{\omega}^t)$.

\emph{Proof:} The Lagrangian function of (\ref{dual SDP}) is defined as 
\begin{align}\label{Lag fun}
&f(\mathcal{Q},\mathbf{P}^r,\mathbf{P}^t,\bm{\Lambda}^t,\bm{\Lambda}^r,\mathbf{v},\bm{\alpha},\lbrace \mathbf{\bar{Q}}_p,\bm{\Lambda}_p, \mathbf{U}_p \ | \ p \in \mathcal{C}_P \rbrace) \nonumber \\
= & \sum_{p=1}^P \Bigl( <\mathbf{Q}_p,\mathbf{Y}_p>_{\mathbb{R}} -  <\begin{bmatrix} P\bm{\Lambda}^r  & \bm{\Lambda}_p  \\ (\bm{\Lambda}_p)^H & \bm{\Lambda}^t \end{bmatrix}, \begin{bmatrix} \mathbf{P}^r / P  & \mathbf{\bar{Q}}_p  \\ \mathbf{\bar{Q}}_p^H & \mathbf{P}^t \end{bmatrix}>_{\mathbb{R}} \nonumber \\
& \  - <\mathbf{U}_p,\mathbf{\bar{Q}}_p - R(\mathbf{Q}_p)>_{\mathbb{R}} \Bigr)- \sum_{k=0}^{N_R-1}\mathbf{v}_{(k+1)}(\delta(k) \nonumber \\
& -\sum_{i=1}^{N_R-k}\mathbf{P}^r_{(i,i+k)}) -\sum_{k=0}^{N_T-1}\bm{\alpha}_{(k+1)} \bigr(\delta(k)-\sum_{i=1}^{N_T-k}\mathbf{P}^t_{(i,i+k)} \bigl)\nonumber \\
\overset{(\text{a})}=  & \sum_{p=1}^P \Bigl( <\mathbf{Q}_p,\mathbf{Y}_p>_{\mathbb{R}} - < \mathbf{P}^r, \bm{\Lambda}^r>_{\mathbb{R}} - < \mathbf{P}^t, \bm{\Lambda}^t>_{\mathbb{R}} \nonumber \\
& \ - <2\bm{\Lambda}_p, \mathbf{\bar{Q}}_p>_{\mathbb{R}} + <\mathbf{Q}_p,R^*(\mathbf{U}_p)>_{\mathbb{R}} - 
 <\mathbf{U}_p,\mathbf{\bar{Q}}_p>_{\mathbb{R}} \Bigr) \nonumber \\
& + <\mathbf{P}^r,\mathbf{T}^r>_{\mathbb{R}} + <\mathbf{P}^t,\mathbf{T}^t>_{\mathbb{R}} - \sum_{k=0}^{N_R-1}\mathbf{v}_{(k+1)} \delta(k)   \nonumber \\
& - \sum_{k=0}^{N_T-1}\bm{\alpha}_{(k+1)} \delta(k)\nonumber \\
\overset{(\text{b})}=  & <\mathbf{P}^r,\mathbf{T}^r-P\bm{\Lambda}^r >_{\mathbb{R}} - \sum_{p=1}^P  <2\bm{\Lambda}_p + \mathbf{U}_p,\mathbf{\bar{Q}}_p>_{\mathbb{R}}  \nonumber \\
&  + <\mathbf{P}^t,\mathbf{T}^t-P\bm{\Lambda}^t >_{\mathbb{R}} +  \sum_{p=1}^P  <\mathbf{Q}_p,\mathbf{Y}_p +R^*(\mathbf{U}_p)>_{\mathbb{R}} \nonumber \\
&   -\bm{\alpha}_{(1)}-\mathbf{v}_{(1)},
\end{align}
where $\bm{\Lambda}_p,\bm{\Lambda}^t, \bm{\Lambda}^r, \mathbf{v},\bm{\alpha} $ denote the Lagrangian multiplier. The equation $(a)$ holds because of the following simplifications:
\begin{equation}
\begin{aligned}
<\mathbf{P}^r,\mathbf{T}^r>_{\mathbb{R}} &= \sum_{k=0}^{N_R-1}\mathbf{v}_{(k+1)}\sum_{i=1}^{N_R-k}\mathbf{P}^r_{(i,i+k)}, \\
\ <\mathbf{P}^t,\mathbf{T}^t>_{\mathbb{R}} &= \sum_{k=0}^{N_T-1}\bm{\alpha}_{(k+1)}\sum_{i=1}^{N_T-k}\mathbf{P}^t_{(i,i+k)},
\end{aligned}
\end{equation}
where $\mathbf{T}^r$ and $\mathbf{T}^t$ are both Toeplitz matrices, which are respectively defined as 
\begin{equation}
\begin{aligned}
\mathbf{T}^r_{(i,j)} =
\begin{cases}
\mathbf{v}_{(j-i)}, & j - i \geq 0, \\
\mathbf{v}_{(i-j)}^{*}, & j - i < 0, \\
\end{cases} \nonumber \\
\mathbf{T}^t_{(i,j)} =
\begin{cases}
\bm{\alpha}_{(j-i)}, & j - i \geq 0, \\
\bm{\alpha}_{(i-j)}^{*}, & j - i < 0. \\
\end{cases}
\end{aligned} 
\end{equation}
The equation $(b)$ holds because
\begin{equation}
<\mathbf{U}_p, R(\mathbf{Q}_p)>_{\mathbb{R}}=<\mathbf{Q}_p, R^*(\mathbf{U}_p)>_{\mathbb{R}}.
\end{equation}

Moreover, the dual matrix $\bm{\Gamma}_p = \begin{bmatrix} P\bm{\Lambda}^r  & \bm{\Lambda}_p  \\ (\bm{\Lambda}_p)^H & \bm{\Lambda}^t \end{bmatrix}$ needs to be a positive semidefinite (PSD) matrix to ensure that the lower bound of the primal problem is larger than the optimal value of the dual problem. 

Based on the above discussion, we have the dual problem of (\ref{dual SDP}), i.e.,
\begin{align}\label{Lag fun}
\underset{\mathcal{Q},\mathbf{P}^r,\mathbf{P}^t,\lbrace \mathbf{\bar{Q}}_p \ | \ p \in \mathcal{C}_P \rbrace}{\text{inf}} & f(\mathcal{Q},\mathbf{P}^r,\mathbf{P}^t,\bm{\Lambda}^t, \bm{\Lambda}^r,\mathbf{v},\bm{\alpha}, \nonumber \\
& \ \ \ \lbrace \mathbf{\bar{Q}}_p,\bm{\Lambda}_p, \mathbf{U}_p \ | \ p \in \mathcal{C}_P \rbrace)  \nonumber \\
\text{s.t.} \ \ \ \ \ \ \ \ \ \ \ \ & \bm{\Gamma}_p = \begin{bmatrix}P\bm{\Lambda}^r  & \bm{\Lambda}_p  \\ (\bm{\Lambda}_p)^H & \bm{\Lambda}^t \end{bmatrix}\succeq 0.
\end{align}

The function $f$ attains its infimum with respect to $\mathbf{Q}_p$, $\mathbf{\bar{Q}}_p$, $\mathbf{P}^r$ and $\mathbf{P}^t$ only when $\mathbf{Y}_p = -R^*(\mathbf{U}_p)$, $\mathbf{U}_p = -2\bm{\Lambda}_p $, $\mathbf{T}^r = P\bm{\Lambda}^r \succeq 0 $ and $\mathbf{T}^t = P\bm{\Lambda}^t \succeq 0 $, respectively. Therefore,  substituting $\mathbf{U}_p = -2\bm{\Lambda}_p $, $2\bm{\Lambda}_p = \mathbf{\bar{Y}}_p$, $\bm{\Lambda}^r = \mathbf{T}^r /P$ and $\bm{\Lambda}^t = \mathbf{T}^t /P$ into the Lagrangian function $f$  yields the following expression,
\begin{align}\label{simplified lag fun}
- \bm{\alpha}_{(1)}-\mathbf{v}_{(1)} = -\frac{1}{N_R} Tr( \mathbf{T}^r ) - \frac{1}{N_T} Tr( \mathbf{T}^t).
\end{align}
Under the constraints $\bm{\Gamma}_p \succeq 0$, $\mathbf{Y}_p = R^*(\mathbf{\bar{Y}}_p)$, $\mathbf{T}^r\succeq 0 $ and $\mathbf{T}^t \succeq 0 $, maximizing (\ref{simplified lag fun}) can be formulated as the following optimization problem
\begin{align}\label{dual primal SDP} 
 & \underset{\mathbf{T}^r,\mathbf{T}^t}{\text{max}} \ \ \ -(\frac{1}{N_R} Tr(\mathbf{T}^r) + \frac{1}{N_T} Tr(\mathbf{T}^t)) \nonumber \\
{\rm s.t.} & \ \ \  \begin{bmatrix}2\mathbf{T}^r/2  & \frac{1}{2}\mathbf{\bar{Y}}_p  \\ \frac{1}{2} \mathbf{\bar{Y}}_p^H & 2\mathbf{T}^t/2P \end{bmatrix} \succeq 0, \mathbf{\bar{Y}}_p = R(\mathbf{Y}_p), p \in \mathcal{C}_P, 
\end{align}
which holds the same expression as (\ref{primal SDP}). Moreover, according to \cite{5,6,7}, the Toeplitz matrices $ \mathbf{T}^r$ and $ \mathbf{T}^t$ with positive semidefinite constraints can be represented as follows:
\begin{align}\label{vand decom}
 \mathbf{T}^r = \mathbf{W}(\bm{\omega}^r)  \mathbf{D}^r \mathbf{W}^H(\bm{\omega}^r), \  \mathbf{T}^t = \mathbf{W}(\bm{\omega}^t)  \mathbf{D}^t \mathbf{W}^H(\bm{\omega}^t),
\end{align}
where $\bm{\omega}^r = [\omega^r_1,\cdots,\omega^r_L]$ and $\bm{\omega}^t = [\omega^t_1,\cdots,\omega^t_L]$ respectively correspond to DOAs $\lbrace \theta_{l} \ | \ l \in \mathcal{C}_L \rbrace$ and DODs $\lbrace \phi_{l} \ | \  \in \mathcal{C}_L \rbrace$.

\section{EFFICIENT ADMM FOR 2D DOA ESTIMATION}
The primal SDP objective (\ref{primal SDP}) resembles those in the classical ANM literature \cite{2,3,4}, where trace minimization is often employed to recover the associated positive semidefinite Toeplitz matrix. It follows that this objective can also be addressed by algorithms designed for conventional ANM problems, with the angle information extracted via (\ref{vand decom}). Accordingly, this section adopts the alternating direction method of multipliers (ADMM), a method widely employed in ANM literature, to solve (\ref{primal SDP}).

The augmented Lagrangian function of (\ref{primal SDP}) is defined as
\begin{align}\label{aug lag fun}
& f_{\rho}(\mathbf{T}^r, \mathbf{T}^t, \lbrace \mathbf{Z}_p, \bm{\Psi}_p \ | \ p \in \mathcal{C}_P \rbrace ) \nonumber \\
= & \sum_{p=1}^P  <\bm{\Psi}_p,\mathbf{Z}_p - \begin{bmatrix}2\mathbf{T}^r  & \mathbf{\bar{Y}}_p  \\ \mathbf{\bar{Y}}_p^H & 2\mathbf{T}^t/P \end{bmatrix}> + \frac{Tr(\mathbf{T}^r)}{N_r} + \frac{ Tr(\mathbf{T}^t)}{N_t} \nonumber \\
&  + \frac{\rho}{2} \sum_{p=1}^P  \Bigg | \Bigg | \mathbf{Z}_p - \begin{bmatrix}2\mathbf{T}^r  & \mathbf{\bar{Y}}_p  \\ \mathbf{\bar{Y}}_p^H & 2\mathbf{T}^t/P \end{bmatrix} \Bigg | \Bigg |_F^2,
\end{align}
where we define 
\begin{align}\label{aug lag multiplier}
\mathbf{Z}_p = \begin{bmatrix} \mathbf{Z}^r  & \mathbf{\bar{Y}}_p  \\ (\mathbf{\bar{Y}}_p)^H & \mathbf{Z}^t \end{bmatrix}, \bm{\Psi}_p = \begin{bmatrix} \bm{\Psi}^r  & \bm{\hat{\Psi}}_p  \\ (\bm{\hat{\Psi}}_p)^H & \bm{\Psi}^t \end{bmatrix}.
\end{align}
The $g+1$-th iteration of the ADMM comprises the following steps:
\begin{align}\label{T0}
\mathbf{T}^{r,g+1} & = \underset{\mathbf{T}^{r}}{\text{min}} \ f_{\rho}(\mathbf{T}^r, \mathbf{T}^{t,g}, \lbrace \mathbf{Z}_p^{g}, \bm{\Psi}_p^{g} \ | \ p \in \mathcal{C}_P \rbrace ) \nonumber \\
& = \frac{1}{2}\sum_{p=1}^P ( \mathbf{Z}^{r,g} + \bm{\Psi}^r/\rho ) - \frac{P}{2\rho N_R}\mathbf{I}_{N_R},
\end{align}
\begin{align}\label{T1}
 \mathbf{T}^{t,g+1} & = \underset{\mathbf{T}^{t}}{\text{min}} \ f_{\rho}(\mathbf{T}^{r,g+1}, \mathbf{T}^{t}, \lbrace \mathbf{Z}_p^{g}, \bm{\Psi}_p^{g} \ | \ p \in \mathcal{C}_P \rbrace )
\nonumber \\
& = \frac{1}{2}\sum_{p=1}^P ( \mathbf{Z}^{t,g} + \bm{\Psi}^t/\rho ) - \frac{P}{2\rho N_R}\mathbf{I}_{N_R}, 
\end{align}
\begin{align}\label{Zp}
\mathbf{Z}_p^{g+1} & = \underset{\mathbf{Z}_p}{\text{min}} \  f_{\rho}(\mathbf{T}^{r,g+1}, \mathbf{T}^{t,g+1}, \lbrace \mathbf{Z}_p, \bm{\Psi}_p^{g} \ | \ p \in \mathcal{C}_P \rbrace ) \nonumber \\
& = \Xi(\begin{bmatrix} 2\mathbf{T}^{r,g+1}  & \mathbf{\bar{Y}}_p  \\ (\mathbf{\bar{Y}}_p)^H & 2\mathbf{T}^{t,g+1}/P \end{bmatrix} - \bm{\Psi}^g_p / \rho), 
\end{align}
\begin{align}\label{Psip}
\bm{\Psi}_p^{g+1} & = \underset{\bm{\Psi}_p}{\text{min}} \  f_{\rho}(\mathbf{T}^{r,g+1}, \mathbf{T}^{t,g+1}, \lbrace \mathbf{Z}_p^{g+1}, \bm{\Psi}_p \ | \ p \in \mathcal{C}_P \rbrace ) \nonumber \\
& = \bm{\Psi}_p^{g} + \rho (\mathbf{Z}_p^{g+1} - \begin{bmatrix} 2\mathbf{T}^{r,g+1}  & \mathbf{\bar{Y}}_p  \\ (\mathbf{\bar{Y}}_p)^H & 2\mathbf{T}^{t,g+1}/P \end{bmatrix}),
\end{align}
where $\rho$ is a penalty parameter, the operator $\Xi(\cdot)$ projects a matrix onto the positive semidefinite cone, which is implemented by performing eigenvalue decomposition and retaining only the non-negative eigenvalues. Once the estimates $\mathbf{\tilde{T}}^{r}$ and $\mathbf{\tilde{T}}^{t}$ are obtained, the DOAs and DODs can be extracted via the Vandermonde decomposition (\ref{vand decom}).

The overall algorithm is summarized in algorithm \ref{alg:lc ALM}

\appendices
\section{the invertibility of $\mathbf{E}_p$}
Before we discuss the invertibility, some useful properties about $K_p(\omega^r)$ are given first (which also hold for $K_p(\omega^t)$), i.e.,
\begin{align}\label{properties about K}
K_p(\omega^r) &= \frac{1}{p}K_1(p\omega^r), K_p^{'}(\omega^r) = K_1^{'}(p\omega^r), \nonumber \\
K_p^{''}(\omega^r) &= pK_1^{''}(p\omega^r).
\end{align}

According to \cite{2}, $\mathbf{E}$ is invertible if $\mathbf{C}= \begin{bmatrix} 
 \mathbf{E}^{(2,0)}_{p} \mathbf{E}^{(1,1)}_{p} \\
 \mathbf{E}^{(1,1)}_{p} \mathbf{E}^{(0,2)}_{p}
\end{bmatrix} $ and its Schur complement $\mathbf{S}_3$ are both invertible.
Therefore, we begin by proving the invertibility of $\mathbf{C}$.

Similarly, the invertibility of $\mathbf{C}$ relies on $\mathbf{E}_p^{(0,2)}$ and $\mathbf{S}_1$. Therefore, we first discuss the invertibility of $\mathbf{E}_p^{(0,2)}$ by assuming $f_c = \max \lbrace 2M_a,2N_a \rbrace \geq 128$. According to (\ref{properties about K}) and \cite{2}, $K_p^{(0,2)}(\lbrace 0,0\rbrace) = K_1(0)K_1^{''}(0)=K_1^{(0,2)}(\lbrace 0,0\rbrace)=\frac{-\pi^2f_c(f_c+4)}{3}$, and the $(m,n)$-th entry $\mathbf{E}_{p,(m,n)}^{(0,2)}= K_p(\omega_m^r - \omega_n^r) K_p^{''}(\omega_m^t - \omega_n^t)=K_1(p(\omega_m^r - \omega_n^r)) K_1^{''}(p(\omega_m^t - \omega_n^t))$.  Thus $|| \ |K_p^{(0,2)}(\lbrace 0,0\rbrace)|\mathbf{I}_{L\times L} - \mathbf{E}_p^{(0,2)}||_{\infty}$ = $||\ |K_1^{(0,2)}(\lbrace 0,0\rbrace)|\mathbf{I}_{L\times L} - \mathbf{E}_1^{(0,2)}||_{\infty} \leq 0.3539f_c^2$. Moreover, according to \cite{2}, the symmetric matrix $\mathbf{E}_p^{(0,2)}$ is invertible if  $||\mathbf{I}_{L\times L} - \mathbf{E}_p^{(0,2)}||_{\infty} < 1$, thus $||\mathbf{I}_{L\times L} - \frac{ \mathbf{E}_p^{(0,2)}}{|K_p^{(0,2)}(\lbrace 0,0\rbrace)|}||_{\infty} = \frac{||\ |K_p^{(0,2)}(\lbrace 0,0\rbrace)|\mathbf{I}_{L\times L} - \mathbf{E}_p^{(0,2)}||_{\infty}}{|K_p^{(0,2)}(\lbrace 0,0\rbrace)|} =\frac{1.0608f_c^2}{\pi^2f_c(f_c+4)}<1$ proves the invertibility of $ \mathbf{E}_p^{(0,2)}$.

Next, we consider the invertibility of $\mathbf{S}_1$. Notice that as $ ||\ |K_p^{(0,2)}(\lbrace 0,0\rbrace)|\mathbf{I}_{L\times L} - \mathbf{E}_p^{(2,0)} ||_{\infty} =  ||\ |K_p^{(0,2)}(\lbrace 0,0\rbrace)|\mathbf{I}_{L\times L} - \mathbf{E}_p^{(0,2)} ||_{\infty}$ due to the definition of Fejér kernel, we have $|| \ |K_p^{(0,2)}(\lbrace 0,0\rbrace)|\mathbf{I}_{L\times L} - \mathbf{S}_1||_{\infty} \leq ||\ |K_p^{(0,2)}(\lbrace 0,0\rbrace)|\mathbf{I}_{L\times L} - \mathbf{E}_p^{(0,2)} ||_{\infty} + ||\mathbf{E}_p^{(1,1)} ||_{\infty}^2 ||(\mathbf{E}_p^{(0,2)})^{-1}||_{\infty}$. Meanwhile, as the  $(m,n)$-th entry  $\mathbf{E}_{p,(m,n)}^{(1,1)} = K_1^{'}(p(\omega_m^r - \omega_n^r)) K_1^{'}(p(\omega_m^t - \omega_n^t))$, the upper bound of $||\mathbf{E}_{p}^{(1,1)}||_{\infty}$ is the same as the upper bound of $||\mathbf{E}_{1}^{(1,1)}||_{\infty}$ according to \cite{1,2}, i.e., $0.1576f_c^2$. Similarly, the upper bound of $||(\mathbf{E}_p^{(0,2)})^{-1}||_{\infty}$ is $ \frac{0.3399}{f_c^2}$. Thus, as  $|K_p^{(0,2)}(\lbrace 0,0\rbrace)| = \frac{\pi^2f_c(f_c+4)}{3} \geq \frac{\pi^2f_c^2}{3} + \frac{4\pi^2f_c^2}{3\times 128} = \frac{11\pi^2f_c^2}{32}$, we can deduce that $ \frac{|| \ |K_p^{(0,2)}(\lbrace 0,0\rbrace)|\mathbf{I}_{L\times L} - \mathbf{S}_1||_{\infty}}{|K_p^{(0,2)}(\lbrace 0,0\rbrace)|} \leq 1$, which proves the invertibility of $\mathbf{S}_1$.

According to the above discussion, $\mathbf{C}$ is invertible. Then, we further discuss the invertibility of $\mathbf{S}_3$ from the following inequality,
\begin{align}\label{IS3}
|| \mathbf{I}_{L \times L} - \mathbf{S}_3 ||_{\infty} \leq &|| \mathbf{I}_{L \times L} - \mathbf{E}^{(0,0)}_p ||_{\infty} + ||\mathbf{S}_2 ||_{\infty}^2 ||\mathbf{S}_1^{-1}||_{\infty} + \nonumber \\
&|| \mathbf{E}^{(0,1)}_p ||_{\infty}^2 ||(\mathbf{E}^{(0,2)}_p)^{-1}||_{\infty}.
\end{align}
Based on (\ref{properties about K}) and \cite{1}, we list the upper bounds of the terms in (\ref{IS3}).
\begin{align}
&|| \mathbf{E}^{(0,1)}_p ||_{\infty} \leq \frac{7.723\times 10^{-2}}{p}, \\
&||  \mathbf{I}_{L \times L} - \mathbf{E}^{(0,0)}_p ||_{\infty} \nonumber \\
& \ \ \ \ \leq ||  \mathbf{I}_{L \times L} - \frac{1}{p^2} \mathbf{I}_{L \times L}||_{\infty} + \frac{1}{p^2} || \mathbf{I}_{L \times L} - \mathbf{E}^{(0,0)}_1||_{\infty} \nonumber \\
&\ \ \ \ = 1 + \frac{4.854\times 10^{-2}-1}{p^2}, \\
& ||\mathbf{S}_2||_{\infty} \leq || \mathbf{E}_p^{(1,0)}||_{\infty} + || \mathbf{E}_p^{(1,1)}||_{\infty}  || \mathbf{E}_p^{(0,1)}||_{\infty} ||(\mathbf{E}_p^{(0,2)})^{-1}||_{\infty} \nonumber \\
&\ \ \ \ \ \ \ \ \ = \frac{8.1369\times 10^{-2}f_c}{p}, \\
&  ||\mathbf{S}_1^{-1}||_{\infty} \leq \frac{0.3408 / f_c^2}{ |K_p^{(0,2)}(\lbrace 0,0\rbrace)| - || \ |K_p^{(0,2)}(\lbrace 0,0\rbrace)|\mathbf{I}_{L\times L} - \mathbf{S}_1||_{\infty}}, \nonumber \\
&\ \ \ \ \ \ \ \ \ \ \ = \frac{0.1115}{f_c^4}.
\end{align}
Therefore, we have 
\begin{align}\label{IS3}
|| \mathbf{I}_{L \times L} - \mathbf{S}_3 ||_{\infty} & \leq \frac{p^2-1+5.0567\times 10^{-2}}{p^2} + \frac{7.3830\times 10^{-4}}{f_c^2p^2} \nonumber \\
& \approx \frac{p^2-1+5.0567\times 10^{-2}}{p^2} \le 1,
\end{align}
which proves the invertibility of $\mathbf{S}_3$. Combining the above results, we conclude that $\mathbf{E}$ is invertible.
\section{Proof of Lemma I}
According to (\ref{polynomial coeff}), we can obtain the upper bound of the interpolation coefficients.
\begin{align}\label{coeff infty upper}
& ||\bm{\alpha}_p||_{\infty} = || \mathbf{S}^{-1}_{3} \begin{bmatrix} \mathbf{d}^{*}_{l,(p)} \\  \cdots \\
\mathbf{d}^{*}_{L,(p)}
\end{bmatrix} ||_{\infty} \leq \frac{1}{\sqrt{P}}|| \mathbf{S}^{-1}_{3}||_{\infty}, \\
& ||\bm{\beta}_p||_{\infty} = ||\mathbf{S}_1^{-1} \mathbf{S}_2 \mathbf{S}_3^{-1}  \begin{bmatrix} \mathbf{d}^{*}_{l,(p)}\\  \cdots \\
\mathbf{d}^{*}_{L,(p)}
\end{bmatrix} ||_{\infty} \leq \frac{1}{\sqrt{P}}|| \mathbf{S}_1^{-1} \mathbf{S}_2 \mathbf{S}_3^{-1} ||_{\infty}, \\
& ||\bm{\epsilon}_p||_{\infty} = || (\mathbf{E}_p^{(0,2)})^{-1} (\mathbf{E}_p^{(1,1)}\mathbf{S}_1^{-1} \mathbf{S}_2 - \mathbf{E}_p^{(0,1)}) \\
& \ \ \ \ \ \ \ \ \ \ \ \ \ \ \ \mathbf{S}^{-1}_3 \begin{bmatrix} \mathbf{d}^{*}_{l,(p)} \\  \cdots \\
\mathbf{d}^{*}_{L,(p)}\end{bmatrix} ||_{\infty} \\
& \ \ \ \ \ \ \ \ \leq \frac{1}{\sqrt{P}}|| (\mathbf{E}_p^{(0,2)})^{-1} (\mathbf{E}_p^{(1,1)}\mathbf{S}_1^{-1} \mathbf{S}_2 - \mathbf{E}_p^{(0,1)})\mathbf{S}^{-1}_3 ||_{\infty}.
\end{align}
We begin by illustrating the upper bound of $||\bm{\alpha}_p||_{\infty}$.  Since the invertibility of $\mathbf{S}_{3}$ has been proved in Appendix A, thus we have \cite{1}
\begin{align}\label{alpha upper bound}
\frac{1}{\sqrt{P}}||\mathbf{S}_3^{-1}||_{\infty} \leq \frac{1}{(1-|| \mathbf{I}_{L \times L} - \mathbf{S}_3  ||_{\infty})\sqrt{P}} \leq \frac{1.0533p^2}{\sqrt{P}}.
\end{align}

Secondly, based on the bounds provided in Appendix A, it is straightforward to compute the upper bound of $||\bm{\beta}_p||_{\infty}$, i.e.,
\begin{align}\label{beta upper bound}
||\bm{\beta}_p||_{\infty} \leq \frac{1}{\sqrt{P}} ||\mathbf{S}_1^{-1}||_{\infty} ||\mathbf{S}_2||_{\infty} || \mathbf{S}_3^{-1}||_{\infty} = \frac{0.9556\times 10^{-2}p}{\sqrt{P}f_c^3}.
\end{align}

Finally, the upper bound of $||\bm{\epsilon}_p||_{\infty}$ can be simply computed based on the results in Appendix A and Appendix B, i.e.,
\begin{align}\label{epsilon upper bound}
||\bm{\epsilon}_p||_{\infty} \leq \ & ||(\mathbf{E}_p^{(0,2)})^{-1}||_{\infty} || \mathbf{E}_p^{(1,1)} ||_{\infty} ||\mathbf{S}_1^{-1}||_{\infty} || \mathbf{S}_2 ||_{\infty} ||\mathbf{S}^{-1}_3 ||_{\infty} \nonumber \\
& + ||(\mathbf{E}_p^{(0,2)})^{-1}||_{\infty} ||\mathbf{E}_p^{(0,1)}||_{\infty} ||\mathbf{S}^{-1}_3 ||_{\infty} \nonumber \\
& \!\!\!\!\!\!\!\!  = \frac{0.0512\times 10^{-2}p}{f_c^3} + \frac{2.765\times 10^{-2}p}{f_c} \nonumber \\
 & \!\!\!\!\!\!\!\! \approx \frac{2.765\times 10^{-2}p}{f_c}.
\end{align}

\section{Proof of Proposition I}
Based on \cite{6,4}, the following proof indicates that the constraint  $|| \mathcal{Q} ||_{\mathcal{A}}^{*} < 1$ in (\ref{dual problem}) is equivalent to the PSD constraint in  (\ref{atomic norm min}). Specifically, we first show that $|| \mathcal{Q} ||_{\mathcal{A}}^{*} < 1$ holds only if a polynomial $V(\phi , \theta) > 0$ holds, then we prove that $V(\phi , \theta) > 0$ leads to PSD constratints in (\ref{atomic norm min}). Finally, we indicate that PSD constraints also result in $V(\phi , \theta) > 0$.

According to the definition of dual norm in (\ref{dual norm trace}), i.e.,
\begin{align}\label{dual norm trace}
|| \mathcal{Q} ||_{\mathcal{A}}^{*} = \underset{ \phi , \theta , \bar{\beta}^{p}}{\text{sup}} \sum_{p=1}^{P} RTr(\mathbf{b}^T(\phi,p)\mathbf{Q}_p^H \mathbf{a}(\theta,p) \bar{\beta}^{p}),
\end{align}
the array steering vectors $\mathbf{a}(\theta,p)$ and $\mathbf{b}(\phi,p)$ vary with the subcarrier index. Thus, we leverage the irregular Vandermonde vectors $\mathbf{w}^r \in \mathbb{C}^{N_R}$ and $\mathbf{w}^t\in \mathbb{C}^{N_T}$ defined in \emph{Definition 1} to respectively aggregate the steering vectors $\mathbf{a}(\theta,p), \mathbf{b}(\phi,p), p = 1,\cdots,P$. Meanwhile, to construct a homogeneous representation of $\mathbf{Q}_p$, we define $\mathbf{\bar{Q}}_p = R(\mathbf{Q}_p)$, where $R(\cdot)$ is given in (\ref{R trans}).  Then, we can provide a compact dual polynomial vector based on $\mathcal{Q}$, i.e.,
\begin{align}\label{dual polynomial vector}
\bm{\chi}_{\phi , \theta} &= [ \mathbf{b}^T(\phi,1)\mathbf{Q}_1^H\mathbf{a}(\theta,1),\cdots, \mathbf{b}^T(\phi,P)\mathbf{Q}_1^H\mathbf{a}(\theta,P)] \nonumber \\
& = [(\mathbf{w}^t)^T\mathbf{\bar{Q}}_1^H \mathbf{w}^r,\cdots,(\mathbf{w}^t)^T\mathbf{\bar{Q}}_P^H \mathbf{w}^r].
\end{align}
The dual norm becomes :
\begin{align}
|| \mathcal{Q} ||_{\mathcal{A}}^{*}  = \underset{ \phi , \theta , \bar{\beta}^{p}}{\text{sup}}  RTr(\bm{\chi}_{\phi , \theta} \bar{\bm{\beta}}) = \underset{ \phi , \theta}{\text{sup}} || \bm{\chi}_{\phi , \theta} ||_2,
\end{align}
where $\bar{\bm{\beta}} = [\bar{\beta}^{1},\cdots,\bar{\beta}^{P}]^T$. The second equation is obtained by Cauchy Inequality and the assumption that $|| \bar{\bm{\beta}}|| < 1$. 

Therefore, the constraint $|| \mathcal{Q} ||_{\mathcal{A}}^{*} \leq 1$ in (\ref{dual problem}) holds only if $V(\phi , \theta) > 0$ holds, where $V(\phi , \theta)$ denotes
\begin{align}
V(\phi , \theta) & = 1 -  || \bm{\chi}_{\phi , \theta} ||_2^2 = 1- Tr(\bm{\chi}_{\phi , \theta}^H \bm{\chi}_{\phi , \theta}) \\
& = 1 -  \sum_{p=1}^{P} (\mathbf{w}^r)^H \mathbf{\bar{Q}}_p (\mathbf{w}^t)^* (\mathbf{w}^t)^T\mathbf{\bar{Q}}_p^H  \mathbf{w}^r. \nonumber
\end{align}

Next, we argue that $V(\phi , \theta) > 0$ leads to PSD constraints in (\ref{dual SDP}) under the condition $|\mathbf{b}^T(\phi,p)\mathbf{Q}_p^H \mathbf{a}(\theta,p)|<\frac{1}{\sqrt{P}}, p\in \mathcal{C}_P$. This part of proof includes three steps: firstly, we need to show that $V(\phi , \theta) > 0$ leads to two PSD matrices, i.e.,
\begin{align}\label{P0 define}
\mathbf{P}^r \succeq 0 \in \mathbb{C}^{N_R\times N_R},  \sum_{i=1}^{N_R-k} \mathbf{P}^r_{(i,i+k)} = \delta(k), 
\end{align}
\vspace{-10pt}
\begin{align}\label{P1 define}
 \mathbf{P}^t \succeq 0 \in \mathbb{C}^{N_T\times N_T},  \sum_{i=1}^{N_T-k} \mathbf{P}^t_{(i,i+k)} = \delta(k).
\end{align}
Then, we need to prove that $\mathbf{P}^r \otimes \mathbf{P}^t$ can construct the PSD matrix $\mathbf{K}_p$, i.e.,
\begin{align}
\mathbf{K}_p = \begin{bmatrix} \frac{1}{P}\mathbf{P}^t \otimes \mathbf{P}^r & vec(\mathbf{\bar{Q}}_p)  \\ vec(\mathbf{\bar{Q}}_p)^H & 1 \end{bmatrix} \succeq 0, \ \ \forall p \in \mathcal{C}_P.
\end{align}
Finally, we conclude that the existence of $\mathbf{K}_p$ definitely results in a PSD matrix $\mathbf{G}_p$ in (\ref{dual SDP}).

To begin with, $V(\phi , \theta) > 0$ implies that the univariate trigonometric polynomials in Lemma III satisfy
\begin{align}
 T_{\mathbf{\bar{Q}}^r (\mathbf{\bar{Q}}^r)^H}( \mathbf{w}^r, \mathbf{w}^r)<1, \ \ \  T_{\mathbf{\bar{Q}}^t (\mathbf{\bar{Q}}^t)^H}( \mathbf{w}^t, \mathbf{w}^t)<1,
\end{align}
where $\mathbf{\bar{Q}}^r = [\mathbf{\bar{Q}}_1 (\mathbf{w}^t)^*, \cdots, \mathbf{\bar{Q}}_P (\mathbf{w}^t)^*]$ and $\mathbf{\bar{Q}}^t = [\mathbf{\bar{Q}}_1^T (\mathbf{w}^r)^*, \cdots, \mathbf{\bar{Q}}_P^T (\mathbf{w}^r)^*]$. According to Lemma III in the univariate case \cite{8}, the following PSD matrices exist, i.e.,
\begin{align}
\mathbf{U} = \begin{bmatrix} \mathbf{P}^r  & \mathbf{\bar{Q}}^r  \\ (\mathbf{\bar{Q}}^r)^H & \mathbf{I}_{P} \end{bmatrix} \succeq 0 , \text{with} \ \mathbf{P}^r \ \text{in (\ref{P0 define}),} 
\end{align}
\begin{align}
\mathbf{O} = \begin{bmatrix} \mathbf{P}^t  & \mathbf{\bar{Q}}^t  \\ (\mathbf{\bar{Q}}^t)^H & \mathbf{I}_{P} \end{bmatrix} \succeq 0, \text{with} \ \mathbf{P}^t \ \text{in (\ref{P1 define}),}
\end{align}
which completes the proof of the first step.

In the second step, note that as $\mathbf{b}^T(\phi,p)\mathbf{Q}_p^H \mathbf{a}(\theta,p) = (\mathbf{w}^t)^T\mathbf{\bar{Q}}_p^H \mathbf{w}^r$, we get $|(\mathbf{w}^t)^T\mathbf{\bar{Q}}_p^H \mathbf{w}^r |< 1/\sqrt{P},\forall p \in \mathcal{C}_P$ according to the assumption of the proposition. Therefore, the following inequality holds, i.e.,
\begin{align}
(\mathbf{w}^r)^H\mathbf{\bar{Q}}_p (\mathbf{w}^t)^* (\mathbf{w}^t)^T\mathbf{\bar{Q}}_p^H \mathbf{w}^r < \frac{(\mathbf{w}^r)^H\mathbf{P}^r\mathbf{w}^r (\mathbf{w}^t)^H\mathbf{P}^t\mathbf{w}^t}{P},
\end{align}
where $(\mathbf{w}^r)^H\mathbf{P}^r\mathbf{w}^r = (\mathbf{w}^t)^H\mathbf{P}^t\mathbf{w}^t = 1$ according to the definitions of $\mathbf{P}^r, \mathbf{P}^t,\mathbf{w}^r$ and $\mathbf{w}^t$. To facilitate the proof, the above inequality can be transformed to the following form:
\begin{align}\label{P0 P1 vecQp inequ}
& (\mathbf{w}^t \otimes \mathbf{w}^r)^H vec(\mathbf{\bar{Q}}_p) vec(\mathbf{\bar{Q}}_p)^H (\mathbf{w}^t \otimes \mathbf{w}^r) < \nonumber \\
& \ \ \ \ \ \ \ \  \ \ \ \ \ \ \ \ \ \ \ \ \ \ \   (\mathbf{w}^t \otimes \mathbf{w}^r)^H \frac{\mathbf{P}^t \otimes \mathbf{P}^r }{P}(\mathbf{w}^t \otimes \mathbf{w}^r).
\end{align}
According to the Theorem 4.26 in \cite{8}, (\ref{P0 P1 vecQp inequ}) results in the existence of the PSD matrix $\mathbf{K}_p$, which completes the proof of the second step.

In the final step, for arbitrary $\mathbf{a}\in \mathbb{C}^{N_R}$, $\mathbf{b}\in \mathbb{C}^{N_T}$ and $d$, the trigonometric polynomial $T_{\mathbf{K}_p}( \mathbf{q}, \mathbf{q})>0$ holds for the vector $[\mathbf{b}^T \otimes \mathbf{a}^T, d]^T$, i.e.,
\begin{align}
T_{\mathbf{K}_p}&( [\mathbf{b}^T \otimes \mathbf{a}^T, d]^T, [\mathbf{b}^T \otimes \mathbf{a}^T, d]^T) =  \\
& \mathbf{a}^H\mathbf{P}^r\mathbf{a} \mathbf{b}^H\mathbf{P}^t\mathbf{b}/P +  \mathbf{a}^H \mathbf{\bar{Q}}_p \mathbf{b}^* d + d^*\mathbf{b}^T \mathbf{\bar{Q}}_p^H \mathbf{a} + dd^*>0. \nonumber 
\end{align}
Meanwhile, the trigonometric polynomial of $\mathbf{G}_p$ for vector $[\mathbf{a}^T, \mathbf{b}^H]^T$ is presented as
\begin{align}
T_{\mathbf{G}_p}&( [\mathbf{a}^T, \mathbf{b}^H]^T, [\mathbf{a}^T, \mathbf{b}^H]^T) = \\
& \mathbf{a}^H\mathbf{P}^r\mathbf{a}/P + \mathbf{b}^H\mathbf{P}^t\mathbf{b} + \mathbf{a}^H\mathbf{\bar{Q}}_p\mathbf{b}^*  \nonumber + \mathbf{b}^T \mathbf{\bar{Q}}_p^H \mathbf{a}.
\end{align}
Let $d=\mathbf{b}^H\mathbf{P}^t\mathbf{b}$,  we can obtain that $d^H=d^*=d>0$ as both $\mathbf{P}^t$ and $\mathbf{P}^r$ are PSD matrices.  Thus,
$T_{\mathbf{G}_p}( [\mathbf{a}^T, \mathbf{b}^H]^T, [\mathbf{a}^T, \mathbf{b}^H]^T) d = T_{\mathbf{K}_p}( [\mathbf{b}^T \otimes \mathbf{a}^T, d]^T, [\mathbf{b}^T \otimes \mathbf{a}^T, d]^T)>0$ for arbitrary vector $[\mathbf{a}^T, \mathbf{b}^H]^T$. We can conclude that $\mathbf{G}_p \succeq 0$ is also a PSD matrix beacause its trigonometric polynomial of $T_{\mathbf{G}_p}( [\mathbf{a}^T, \mathbf{b}^H]^T, [\mathbf{a}^T, \mathbf{b}^H]^T)>0$ .

Finally, we show that the PSD constraints in (\ref{atomic norm min}) also result in $V(\phi , \theta) > 0$. Firstly, we define the following matrix and its trigonometric polynomial:
\begin{align}\label{sum up poly sum i}
& \mathbf{U}_p = \begin{bmatrix} \mathbf{P}^r/P  & \mathbf{\bar{Q}}_p (\mathbf{w}^t)^*  \\ (\mathbf{w}^t)^T\mathbf{\bar{Q}}_p^P & 1 \end{bmatrix}, \nonumber \\
& T_{\mathbf{U}_p}( [\mathbf{a}^T, d_p]^T, [\mathbf{a}^T, d_p]^T)= \mathbf{a}^H\mathbf{P}^r\mathbf{a}/P + d_{p} d_{p}^* \nonumber \\
&\ \ \ \ \ \ \ \ \ \ \ + \mathbf{a}^H \mathbf{\bar{Q}}_p (\mathbf{w}^t)^*d_{p} + d_{p}^*(\mathbf{w}^t)^T\mathbf{\bar{Q}}_p^H \mathbf{a},
\end{align}
where $\mathbf{a} \in \mathbb{C}^{N_R}$ and $d_p$ represent an arbitrary vector and an arbitrary complex constant, respectively.
Meanwhile, the polynomial for PSD matrix $\mathbf{G}_p$ under the vector $[\mathbf{a}^T,d_p(\mathbf{w}^t)^H]^T$  can be expressed as follows
\begin{align}\label{sum xp poly sum i}
&T_{\mathbf{G}_p}( [\mathbf{a}^T,d_p(\mathbf{w}^t)^H]^T, [\mathbf{a}^T
,d_p(\mathbf{w}^t)^H]^T) \nonumber \\
& \ \ =\mathbf{a}^H\mathbf{P}^r\mathbf{a}/P +  d_p^* d_p +  \mathbf{a}^H \mathbf{\bar{Q}}_P (\mathbf{w}^t)^*d_p + d_p^* (\mathbf{w}^t)^T\mathbf{\bar{Q}}_P^H \mathbf{a},
\end{align}
where we use the property $(\mathbf{w}^t)^T \mathbf{P}^t(\mathbf{w}^t)^* = (\mathbf{w}^t)^H \mathbf{P}^t\mathbf{w}^t = 1$ according to $\mathbf{P}^t$ in (\ref{dual SDP}). Note that (\ref{sum xp poly sum i}) is larger than zero since $\mathbf{G}_p$ is a PSD matrix, thus $\mathbf{U}_p$ is also a PSD matrix as (\ref{sum xp poly sum i}) equals to (\ref{sum up poly sum i}). 

Then, by summing the trigonometric polynomial (\ref{sum up poly sum i}) across $P$ subcarriers, we have
\begin{align}\label{sum up poly sum P}
& \sum_{p=1}^P T_{\mathbf{U}_p}( [\mathbf{a}^T, d_p]^T, [\mathbf{a}^T,  d_p]^T)= \mathbf{a}^H\mathbf{P}^r\mathbf{a} + \sum_{p=1}^P  d_p  d_p^* \nonumber \\
&\ \ \ \ \ \ \ \ \ \ \ + \sum_{p=1}^P \mathbf{a}^H \mathbf{\bar{Q}}_p (\mathbf{w}^t)^* d_p + \sum_{p=1}^P  d_p^*(\mathbf{w}^t)^T\mathbf{\bar{Q}}_p^H \mathbf{a}>0.\nonumber \\
\end{align}
Meanhwile, we can deduce from (\ref{sum up poly sum P}) that
\begin{align}\label{sum up equal u}
\sum_{p=1}^P T_{\mathbf{U}_p}( [\mathbf{a}^T, d_p]^T, [\mathbf{a}^T,  d_p]^T) = T_{\mathbf{U}}( [\mathbf{a}^T, \mathbf{d}^T]^T, [\mathbf{a}^T,  \mathbf{d}^T]^T),
\end{align}
where $\mathbf{d}_{(p)}=d_p$, $\mathbf{U} = \begin{bmatrix} \mathbf{P}^r  & \mathbf{\bar{Q}}^r  \\ (\mathbf{\bar{Q}}^r)^H & \mathbf{I}_{P} \end{bmatrix}$ and $\mathbf{\bar{Q}}^r = [\mathbf{\bar{Q}}_1 (\mathbf{w}^t)^*, \cdots, \mathbf{\bar{Q}}_P (\mathbf{w}^t)^*]$.
Therefore, $\mathbf{U}$ is also a PSD matrix since its trigonometric polynomial is larger than zero.

According to the Schur complement lemma of PSD matrix, we get $\mathbf{P}^r \succeq  \mathbf{\bar{Q}}^r (\mathbf{\bar{Q}}^r)^H$. By multiplying the Schur complement with $(\mathbf{w}^r)^H$ and $\mathbf{w}^r$ on both sides, we get the inequality
\begin{align}
1 > \sum_{p=1}^P (\mathbf{w}^r)^H \mathbf{\bar{Q}}_p (\mathbf{w}^t )^* (\mathbf{w}^t )^T \mathbf{\bar{Q}}_p^H\mathbf{w}^r,
\end{align}
which is exactly the inequality  $V(\phi , \theta) > 0$.
\bibliographystyle{ieeetran}
\bibliography{bare_jrnl}
\end{document}